\pdfoutput=1

\documentclass[11pt]{article}

\usepackage[final]{acl}

\usepackage{times}
\usepackage{latexsym}

\usepackage[T1]{fontenc}

\usepackage[utf8]{inputenc}

\usepackage{microtype}

\usepackage{inconsolata}

\usepackage{graphicx}
\usepackage{makecell} 
\usepackage{type1cm} 
\usepackage{graphicx} 
\usepackage{xspace} 
\usepackage{balance} 
\usepackage{booktabs} 
\usepackage{multirow} 
\usepackage{subcaption} 
\usepackage{bold-extra} 
\usepackage[vlined,linesnumbered,ruled,noend]{algorithm2e} 
\usepackage{microtype} 
\usepackage{siunitx} 
\usepackage{xfrac} 
\usepackage{mathtools} 
\usepackage{url}
\usepackage{natbib}
\usepackage{rotating}
\usepackage{adjustbox}

\PassOptionsToPackage{hyphens}{url} 
\PassOptionsToPackage{bookmarks, pdftex, colorlinks=true, pagebackref=true, backref=page}{hyperref} 
\PassOptionsToPackage{square,numbers}{natbib} 
\usepackage{cleveref} 
\usepackage[hyperpageref]{backref} 
\usepackage{hyphenat} 
\usepackage{ragged2e}
\usepackage{CJKutf8} 

\usepackage[show]{chato-notes} 

\newcommand{\spara}[1]{\smallskip\noindent\textbf{#1}}

\newenvironment{squishlist}
{\begin{list}{$\bullet$}
 {\setlength{\itemsep}{0pt}
     \setlength{\parsep}{3pt}
     \setlength{\topsep}{3pt}
     \setlength{\partopsep}{0pt}
     \setlength{\leftmargin}{1.5em}
     \setlength{\labelwidth}{1em}
     \setlength{\labelsep}{0.5em} } }
{\end{list}}

\renewcommand*\backref[1]{\ifx#1\relax \else (Cited on #1) \fi}

\graphicspath{{images}}

\newcommand{\Cone}{\textbf{\texttt{C1}}\xspace}
\newcommand{\Ctwo}{\textbf{\texttt{C2}}\xspace}
\newcommand{\Cthree}{\textbf{\texttt{C3}}\xspace}

\title{Conspiracy Theories and Where to Find Them on TikTok} 

\author{
\\
\textbf{\textcolor{red}{This manuscript has been accepted to ACL'25}},
\\
\\
 \textbf{Francesco Corso\textsuperscript{1,2}},
 \textbf{Francesco Pierri\textsuperscript{1}},
 \textbf{Gianmarco de Francisci Morales\textsuperscript{2}},
\\
 \textsuperscript{1}Politecnico Di Milano,
 \textsuperscript{2}CENTAI,
\\
 \small{
   \textbf{Correspondence:} \href{mailto:francesco.corso@polimi.it}{francesco.corso@polimi.it}
 }
}

\begin{document}
\maketitle
\begin{abstract}
TikTok has skyrocketed in popularity over recent years, especially among younger audiences. 
However, there are public concerns about the potential of this platform to promote and amplify harmful content.
This study presents the first systematic analysis of conspiracy theories on TikTok. 
By leveraging the official TikTok Research API we collect a longitudinal dataset of $1.5$M videos shared in the U.S. over three years. 
We estimate a lower bound on the prevalence of conspiratorial videos (up to \num{1000} new videos per month) and evaluate the effects of TikTok's Creativity Program for monetization, observing an overall increase in video duration regardless of content.
Lastly, we evaluate the capabilities of state-of-the-art open-weight Large Language Models to identify conspiracy theories from audio transcriptions of videos. 
While these models achieve high precision in detecting harmful content (up to $96\%$), their overall performance remains comparable to fine-tuned traditional models such as RoBERTa.
Our findings suggest that Large Language Models can serve as an effective tool for supporting content moderation strategies aimed at reducing the spread of harmful content on TikTok.
\end{abstract}

\section{Introduction}

Online social platforms have become integral to modern life by facilitating connections, sharing of interests, and the formation of friendships.
Over the past decade, social media have also emerged as the primary source of news and information for the public.\footnote{\href{https://www.pewresearch.org/journalism/fact-sheet/social-media-and-news-fact-sheet}{www.pewresearch.org} accessed on 10/12/2024.}
TikTok, in particular, has become a key platform for online news consumption, with nearly half of TikTok users ($17\%$ of all U.S. adults) regularly relying on it for news and information.\footnote{\href{https://www.pewresearch.org/short-reads/2025/01/17/a-closer-look-at-americans-experiences-with-news-on-tiktok}{www.pewresearch.org} accessed on 30/01/2025.}

TikTok's vast user base and viral content-sharing mechanisms make it a hotbed for malicious content, by facilitating the rapid spread of harmful narratives and dangerous information~\cite{zeng2022content,weimann2023research} while also providing monetary incentives for viral content creators, such as TikTok's Creativity Program.\footnote{\href{https://newsroom.tiktok.com/en-us/unlocking-even-more-opportunities-for-creators-with-the-creativity-program-beta}{TikTok's Creativity Program} accessed on 30/01/2025.}
It is thus essential to ensure a safe environment for users by monitoring and mitigating the impact of malicious actors.
Indeed, researchers have recognized several threats to discourse quality, including disinformation and hate-speech campaigns, political propaganda, and the spread of conspiracy theories~\cite{aimeur2023fake,matamoros2021racism,chaudhari_systematic_2022}.

In this work, we study the presence of online conspiracy theories on TikTok, which are defined as false or unverified narratives that allege secret, often sinister plots or events involving groups, organizations, or governments~\cite{byford_conspiracy_2011}.
The risks of conspiracy theories stem from their potential to warp the perception of reality for users who consume them~\cite{uscinski2017climate}.
Online social media offer a fertile ground for the spread of conspiracies, also driven by the formation of fringe communities~\cite{spohr2017fake}.
These theories can range from relatively harmless speculations to more dangerous beliefs, which fuel misinformation and polarization, and can have dangerous real-world consequences.\footnote{\href{https://www.theguardian.com/us-news/2016/dec/05/washington-pizza-child-sex-ring-fake-news-man-charged}{www.theguardian.com} accessed on 10/12/2024.} 
The rise of Generative AI tools powered by Large Language Models (LLMs) further amplifies concerns about their potential misuse by malicious actors, as these models can generate content that may mislead and harm online users~\cite{augenstein2023factuality}.
At the same time, LLMs offer valuable opportunities for researchers and scientists to analyze large-scale social media data, identify patterns in the spread of misinformation, and understand the dynamics of conspiracy theories~\cite{ziems2024can}.

To date, only a few qualitative approaches have examined conspiracy theories on TikTok, including \citet{weimann2023research}, which explore far-right narratives, and \citet{zenone2022using}, which identify monkeypox-related conspiracy content.
Our work employs a quantitative and systematic approach to data collection and analysis to address the following research questions:
\begin{squishlist}
    \item \textbf{RQ1}: What is the prevalence of videos sharing conspiracy theories on TikTok?
    \item \textbf{RQ2}: Did the Creativity Program impact the supply of conspiratorial content on TikTok?
    \item  \textbf{RQ3}: Can we leverage LLMs to detect conspiracy theories shared on TikTok?
\end{squishlist}
Our contributions are as follows.
To address \textbf{RQ1}, we collect a sample of $1.5$M long videos (with a duration of at least 1 minute) shared on TikTok in the U.S. by approximately $1$M unique users.
Using a data-driven approach and a large-scale dataset of Web documents about conspiracy theories, we identify hashtags that provide strong signals for the presence of conspiratorial videos and estimate a lower bound to their prevalence on TikTok.

To answer \textbf{RQ2}, we measure a small yet significant increase in the average duration of videos shared after the introduction of the Creativity Program, which incentivizes creators to post long videos.
While this finding indicates a shift in the overall content-production behavior on TikTok, we observe no specific effect on conspiratorial videos.

Finally, to address \textbf{RQ3}, we evaluate the ability of state-of-the-art, open-weight LLMs (Llama, Mistral, and Gemma) to detect conspiracy-promoting videos using audio transcriptions extracted via OpenAI Whisper.
We use open-weight models to ensure a fully reproducible and controllable content moderation pipeline, independent of external inference services~\cite{palmer2024using}.
LLMs prove to be highly sensitive to prompt variations \cite{sclar2023quantifying} and, while they outperform our baseline in many configurations, they still exhibit limitations and trade-offs that must be carefully considered for real-world content moderation applications.

\section{Related Work}
\label{sec:related_work}

The study of online conspiracy theories is broad and diverse, with researchers exploring this complex phenomenon from multiple perspectives.
This section reviews relevant studies, emphasizing linguistic analyses, available datasets, and strategies for dealing with this phenomenon.

\citet{douglas_understanding_2019, sutton_conspiracy_2020} adopted a psychological approach to identify key traits of conspiracy theory believers, revealing common factors such as lower levels of income and education. 
\citet{enders_relationship_2023} examined the relationship between social media usage and the tendency to believe in conspiracy theories, finding that social networks alone are insufficient for effectively conveying conspiratorial messages.
Other studies have examined the individuals who share and promote conspiratorial content, tracing their roles back to some of the earliest widely known conspiracy theories~\cite{byford_conspiracy_2011, douglas_what_2023}.



Research has also investigated the linguistic characteristics of conspiracy theories.
\citet{fong2021language} and \citet{meuer_how_2023} analyzed the psychological nuances in the language of conspiracy theories found in online articles and Twitter posts.
Additionally, several datasets have been developed to support researchers in studying conspiracy theories across social media and online-published articles~\cite{pogorelov_wico_2021,phillips_hoaxes_2022,alessandro_miani_loco_2021,golbeck_fake_2018}.

On TikTok, studies have primarily taken a qualitative approach to analyzing harmful content, including conspiracy theories~\cite{zenone2022using}, by manually reviewing videos associated with specific hashtags.
\citet{weimann2023research} identified a consistent presence of far-right extremism in videos, commentary, symbolism, and imagery. Similarly, \citet{basch2021global} examined the top 100 videos related to COVID-19 vaccines and found that many discouraged vaccination.

\section{Methods}
\label{sec:methods}
\subsection{Dataset}

We use the TikTok Research API\footnote{\scriptsize\url{https://developers.tiktok.com/products/research-api}} to collect our dataset.
At the time of writing, this service has severe restrictions on usage and availability.
In particular, academics can request access to a quota of \num{1000} requests per day.

Given the lack of reliable data returned by the API during 2018-2020~\cite{corso2024we}, we run our collection for the period 2021-2023.
This process yields a total of \num{1605696} items, \num{1494831} of which are unique. 
These videos are created by a total of \num{1178303} unique users.

In addition, we use the Research API to collect two random samples of videos ($\sim 10$k videos each) posted in the U.S. one month before and one month after the beginning of the new Creativity Program (May 3, 2023), respectively. 
\Cref{sec:app_data} provides more details on the data collection.

\subsection{Conspiracy Hashtag Enrichment}
Our objective is to identify videos featuring hashtags that explicitly indicate the conspiratorial nature of the content. To achieve this, we manually review the top 30 most frequent seeds from the LOCO conspiracy dataset~\cite{alessandro_miani_loco_2021}, which includes over \num{90000} conspiracy and non-conspiracy articles scraped from the Web.
To minimize ambiguity and reduce noise, we exclude seeds with broader or more general semantics, such as \textit{cancer}, \textit{AIDS}, \textit{coronavirus}, \textit{5G}, and \textit{barackobama}, as these terms often extend beyond conspiratorial contexts.
For instance, the term \textit{cancer} may appear in claims about 'cancer cures being hidden by pharmaceutical companies' but also in discussions of treatment options, awareness campaigns, and personal survivorship stories.
Instead, we focus on the 10 seeds (e.g., \textit{illuminati}, \textit{reptilians}, full list in \Cref{sec:app_llms}) that convey a clearer, more direct association with conspiracy theories to serve as the basis for our labeling and enrichment process.

To identify candidate hashtags associated with conspiracy theories, we start with $\sim\num{650000}$ unique hashtags extracted from the $1.5$M videos in our sample. We then filter out hashtags that appear only once in the dataset, reducing the total to \num{281510}.
For each of these remaining hashtags, we compute the list of co-occurring hashtags and words, along with the number of videos in which the hashtag appears, obtaining the following co-occurrence matrices.
A hashtag/hashtag matrix $H$ with dimensions of \num{281510} $\times$ \num{281510}, and a hashtag/word matrix $W$, with dimensions of \num{281510} $\times$ \num{185973}, where each column represents a unique word in the corpus of video descriptions.

We compute the pairwise similarity between each seed hashtag and the rest of the hashtags, and keep the top \num{20} hashtags by similarity.
We employ the method by \citet{garimella_quantifying_2017} to calculate the pairwise similarity between hashtags:
\begin{equation*}
    \resizebox{.95\hsize}{!}{$sim(h_s,h_t) = \frac{\alpha \cos(W_s,W_t) + (1-\alpha)\cos(H_s,H_t)}{1+\log(df(h_t))}$}
\end{equation*}
The similarity between two hashtags ($h_s$, the seed hashtag, and $h_t$ the target hashtag) is defined as the weighted sum of the cosine similarities between the respective word and hashtag co-occurrence vectors ($W_i$, $H_i$), with the mixing parameter $\alpha$ set to \num{0.3} based on previous work~\cite{garimella_quantifying_2017}.
This sum is discounted by a function of the document frequency of the hashtag $h_t$, to lower the weight of those hashtags that are extremely frequent (e.g., \textit{\#fyp}, \textit{\#viral}).
This process produces a total of $197$ hashtags that can be used to find conspiratorial videos.

\subsection{Evaluation of Conspiracy Hashtags}
\label{subsec:rq1}
We manually validate each of the \num{197} enriched hashtags to identify those actually associated with conspiracy theories.
For each hashtag, we manually inspect a sample of five videos that include it in their description to determine whether they reference conspiracy theories, and thus that the hashtag should be classified as conspiratorial.
This analysis identified different categories of hashtags.
Out of \num{197} collected hashtags, we classify 92 of them as conspiratorial (\textbf{CT}), while we find \num{68} to be noisy or not explicitly related to conspiracy theories (\textbf{NOCT}).
We find \num{28} so-called ``dog whistling'' (\textbf{DW})~\cite{quaranto2022dog} hashtags, i.e., hashtags that seem harmless at first but are used by conspiracy theorists as inconspicuous markers to attract other users interested in the topic being discussed.
During manual validation, a hashtag is classified as ``dog whistling'' if the tagged videos reference conspiracy theories, but the hashtag's literal meaning differs from the context in the videos (e.g., \textit{\#radiowaves} related to weather modification and chemtrails).
There are also a few cases of hashtag hijacking~\cite{hadgu2013political}, in two different flavors.
\textit{Normal}, when conspiracy users use popular hashtags to boost the virality of their content, e.g., \textit{\#skylovers}, \textit{\#billieeilishmusic} (\textbf{HJ}).
And \textit{reverse}, when users add conspiratorial hashtags to disseminate debunking information about a specific conspiracy theory (\textbf{RHJ}).

Given these classes, we use a heuristic to label videos as conspiratorial if they contain at least one hashtag from the classes \textbf{CT} or \textbf{DW}, and exclude videos that have hashtags belonging to the remaining three classes (\textbf{NOCT, HJ, RHJ}).
This process yields \num{1363} conspiracy theory videos.

We further assess the quality of this distant-labeling procedure by manually analyzing a sample of \num{200} random videos, \num{100} from the pool of videos classified as conspiratorial with the hashtag-based distant labeling, and \num{100} from the remaining ones.
Cohen's kappa agreement~\cite{cohen1960coefficient} between our manual labeling validation and the distant labels is $0.81$, which indicates strong agreement.

\subsection{Estimating the Number of U.S. Videos on TikTok}
To address the first research question and establish a lower bound on the number of conspiracy videos on TikTok, we aim to estimate the true size of the population from which our sample was drawn.
We use a simple version of the Good-Turing frequency estimator~\cite{good1953population} to obtain an estimate of the number of long videos on TikTok:
\begin{equation}
    M =\frac{N}{1-\frac{N_1}{K}},
\end{equation}
where $N$ is the number of unique videos in our dataset, $N_1$ is the number of videos that appear only once, and $K$ is the total number of videos in the sample.
We compute this estimate for each month of the collection.
This methodology accounts for the unseen portion of the population and provides a conservative yet reliable estimate of its minimum size, thus ensuring robustness in our analysis.
We also employ a maximum likelihood estimation technique to verify the goodness of the Good-Turing estimator:
\begin{equation}
\label{form:mle}
    M = \frac{N}{1- e^{\frac{-K}{M}}},
\end{equation}
where $N$ is the number of unique videos collected and $K$ is the totality of videos collected.
Since \Cref{form:mle} describes a recursive equation that should converge, we compute \num{1000} iterations of the equation with the starting value of $M$ = \num{10}.
The two estimators produce extremely similar results ($<1\%$ difference, see \Cref{sec:results}).

\subsection{Video Transcripts}
We employ the \texttt{voice\_to\_text} (or \textbf{VTT}) field provided by the API as an input to the LLMs.
The \texttt{voice\_to\_text} is a feature offered by TikTok that allows the user to visualize the dialogue or narration provided by the creator as text on the screen. 
It must be enabled by the content creator at the time of publishing the video.

Of the $\sim1.5$M videos collected, only around $70$k have their transcripts available natively from the API as VTT.
For this reason, we expand the available transcripts by using OpenAI Whisper~\cite{radford2023robust} (\texttt{medium} size model).
To verify the model's performance on our data, we download a set of videos from TikTok for which the \texttt{voice\_to\_text} field is available from the Research API, and use Whisper to extract the transcript.
We then compare the model output both with the retrieved VTT and with a manual transcription of the video (for $\sim 100$ videos).
We use the WER (Word Error Rate) score~\cite{wang2003word} to evaluate the accuracy of the transcriptions.
Both transcriptions (Whisper-generated and API provided) are fairly accurate (median WER $\sim0.15$), which makes OpenAI's Whisper a viable alternative to the native \texttt{voice\_to\_text} field when the latter is unavailable.
We exclude the videos whose transcription is unintelligible (mostly due to having no spoken content or having overbearing music).
Furthermore, we retain only English videos, i.e., those where more than 50\% of their transcript contains English words.
More details about the evaluation are available in \Cref{sec:app_transcr}.

\subsection{Large Language Models and Prompting Strategies}
\label{subsec:prompts}
Our goal is to assess whether open-weight Large Language Models can detect conspiratorial content.
We measure the classification performance of some of the most recent open models: LLama3 (8B)\cite{dubey2024llama}, Mistral (7B)~\cite{jiang2023mistral}, and Gemma (7B)~\cite{team2024gemma}.
We chose these models as they have a comparable number of parameters and size of the context window.
Furthermore, they allow full reproducibility of the results and their computational requirements are modest if compared to larger models.
In addition to evaluating single LLMs as classifiers, we consider an ensemble decision using a majority voting strategy on the outputs of the three models.

We design a zero-shot binary classification task where items about conspiracy theories represent the positive class. Zero-shot prompting implies that the models are not given any example before seeing the actual item to classify.
We do not consider few-shot approaches, as prior literature has shown they do not provide any improvement in this task~\cite{diab2024classifying}.
We use prompts inspired by the ones described by \citet{diab2024classifying}, which were devised to classify Reddit posts from \texttt{r/conspiracy} and other conspiracy-related subreddits.
We use three different variants, as follows:
\begin{squishlist}
    \item \textbf{Simple}: \textit{`Decide whether the given transcription of a video talks about a conspiracy theory or not (if yes output = 1/else  output=0). Provide just your output, no justification.'}
    \item \textbf{With Definition}: We append at the beginning of the simple prompt the definition of conspiracy theory proposed by \citet{douglas_what_2023} (fully listed in \cref{app:definition}).  
    The definition is preceded by the following additional sentence:
    \textit{'Given this definition of conspiracy theory:'}
    \item \textbf{Step-by-step}: we drop the last sentence of the simple prompt (\textit{`Provide just your output, no justification'}) and we append the following text to elicit the chain-of-thought process~\cite{wei2022chain} from the model: \textit{`First, extract the narrative or claim from the text.  Second, decide if the claim talks about a conspiracy theory. Third, answer the question (if yes output = 1/else output=0)'}
\end{squishlist}
Each model thus is given the following input structure: `Transcription:'\{Video Transcription\}. \{Prompt\}.
We describe the input as a `transcription of a video' rather than `textual content of a post' to be faithful to its nature.
We also use numerical binary labels instead of strings to have a `ready-to-use' classification output by the model.
Recall that the ground truth labels for the videos are obtained through the hashtag enrichment process combined with the \emph{distant supervision} approach validated via a manual inspection of $\sim1000$ videos, as described in \Cref{subsec:rq1}.

To test the system in a variety of settings representative of the operational conditions that might be encountered in practice, we use LLMs to classify conspiracy theories videos in three configurations of the data sources available:
\begin{squishlist}
    \item \textbf{\Cone: Balanced dataset with distant labeling:} An ideal case with wide availability of positive-class videos ($\sfrac{1}{2}$ of the dataset);
    \item \textbf{\Ctwo: Unbalanced dataset with distant labeling:} A more realistic setting with a minority of positive examples ($\sfrac{1}{8}$) and an increased representation of the negative class;
    \item \textbf{\Cthree: Unbalanced dataset with manual labeling:} The setting is the same as \Ctwo, but the labels of the positive class are obtained manually.
\end{squishlist}

The balanced dataset \Cone consists of a total of \num{1666} entries, with \num{887} belonging to the positive class (conspiracy) and \num{779} assigned to the negative class (not conspiracy).
The unbalanced datasets contain the same negative-class items as in the balanced setting.
For \Ctwo, the positive class is a random sample of \num{100} positive examples from \Cone.
Instead, in \Cthree the \num{100} examples of the positive class are manually labeled, rather than using the hashtag-based distant supervision.
Specifically, we watch the videos to assess whether they explicitly mention conspiracy theories, e.g., a video with a voice-over (whether a real human voice or a generated one) that describes how NASA lied about the 1969 moon landing.
To account for randomness in the inference process, we run our pipeline with \num{25} different input seeds (see \Cref{sec:app_llms}).
We keep the \texttt{temperature} parameter of the models fixed to \num{0} to enhance reproducibility~\cite{tornberg2024best}.
All these experiments were run on a local machine with a 16-core CPU and a Nvidia A6000 GPU with 48 GB of VRAM.
The time required to run the experiments is approximately $10$ hours. 

\section{Results}
\label{sec:results}


\subsection{Lower Bound on the Number of Conspiracy Videos on TikTok}

\label{sec:RQ1}
\begin{figure}[!t]
    \centering
    \includegraphics[width=0.7\linewidth]{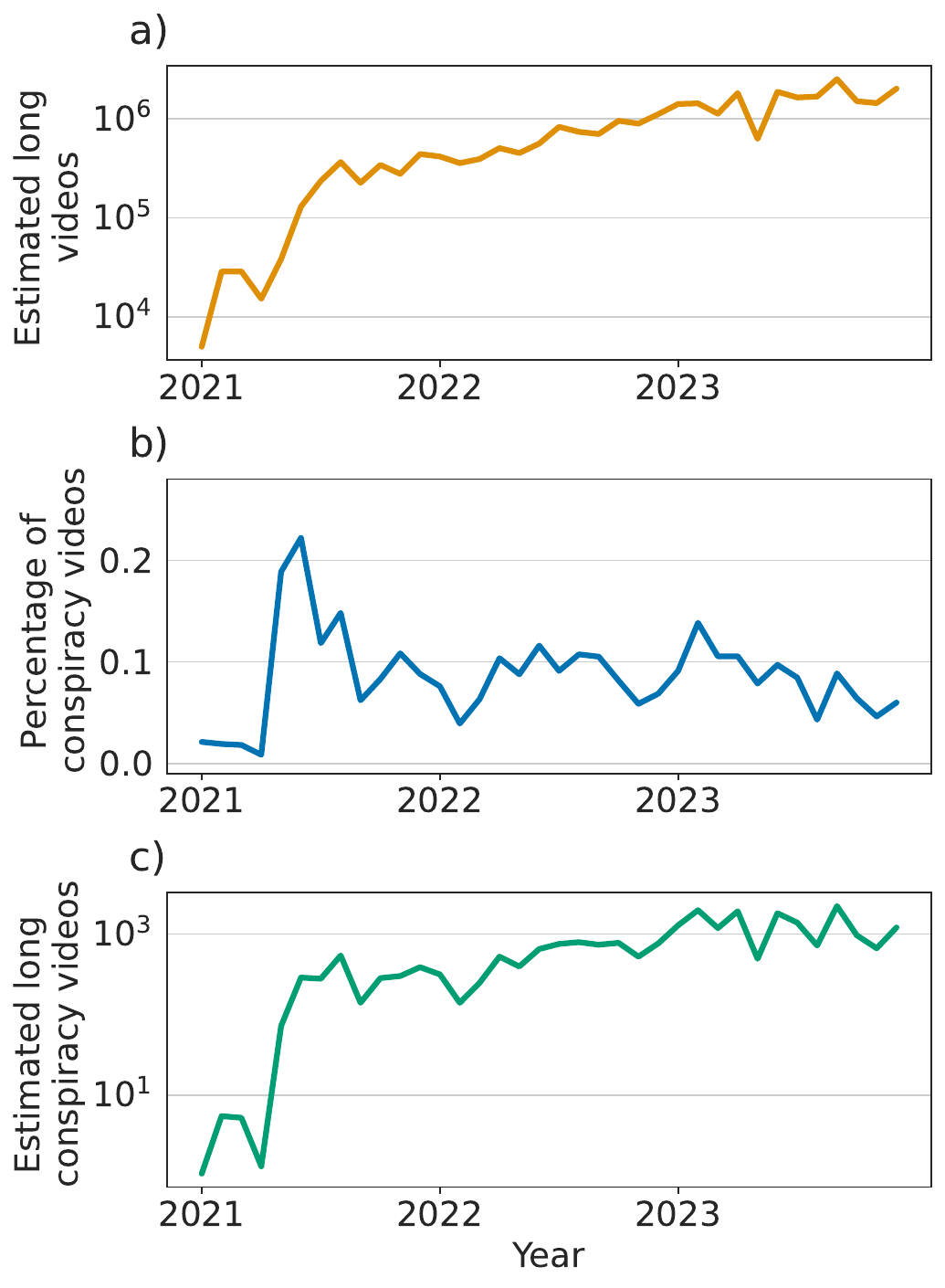}
    \caption{(a) Estimated number of long videos on TikTok U.S. per month. (b) Percentage of conspiracy videos. (c) Estimated number of long conspiracy videos on TikTok U.S.}
    \label{fig:estimate}
\end{figure}

\Cref{fig:estimate} presents the estimated monthly count of long videos shared on TikTok in the U.S., derived via the Good-Turing estimator described in \Cref{sec:methods}.
Over the past three years, the number of videos posted on the platform has grown steadily, increasing by two orders of magnitude.

For each month in our dataset, we compute the number of conspiracy videos and determine the percentage of conspiratorial content using distant labeling.
Combining this information with estimates of the total number of long videos posted on TikTok in the U.S., we can estimate the number of conspiratorial videos.
\nameCrefs{fig:estimate} \labelcref{fig:estimate}b and \labelcref{fig:estimate}c show the temporal trends of these volumes.
In Q3 2021, there was a spike in the upload of conspiracy videos, peaking at $0.2\%$ of our overall sample (N=\num{542}), likely driven by the platform's growth and an increase in the total number of posted videos.
This percentage declines in the following months and stabilizes around $0.1\%$, even as the absolute number of conspiracy videos grows steadily before leveling off, reaching an estimated one thousand conspiracy videos per month in 2023.

Based on these findings, we conclude that conspiracy-related videos are present on TikTok, with a prevalence reaching up to one in every five hundred uploaded videos during specific periods.

\begin{figure}[t]
    \centering
    \includegraphics[width=0.85\linewidth]{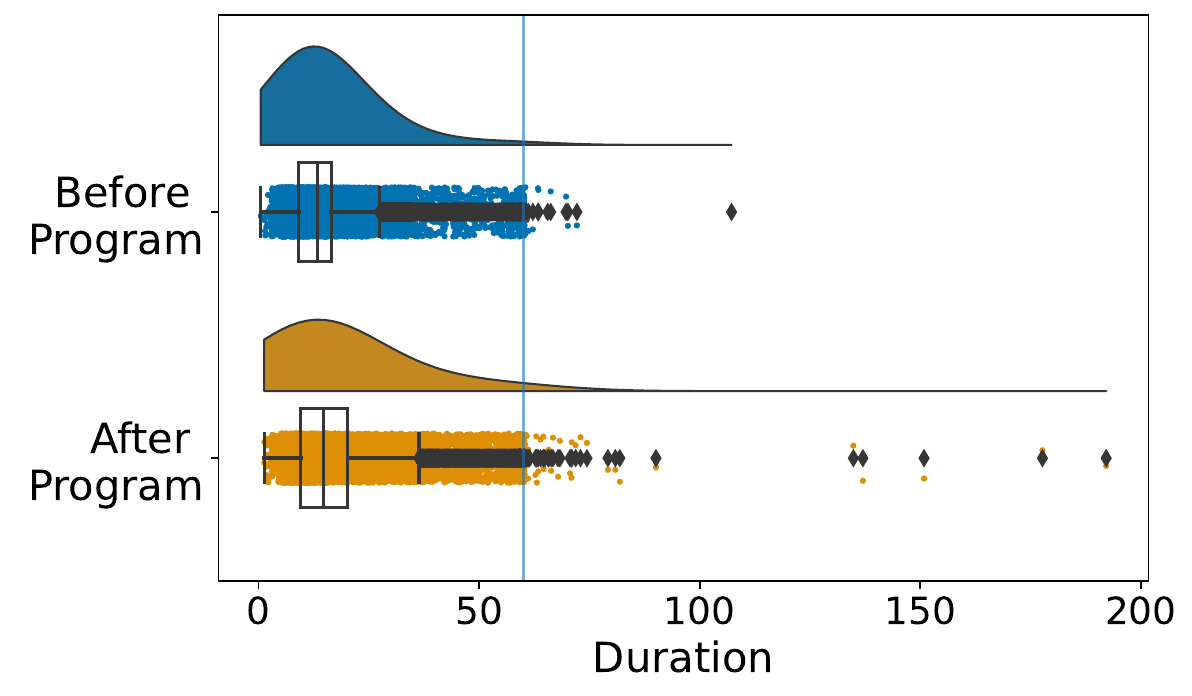}
    \caption{Distribution of the duration in seconds of videos created before (N=\num{42}) and after (N=\num{103}) the beginning of the Creativity Program (May 3, 2023). The vertical blue line indicates $60$ seconds. Medians: before = $13.4$ s, after = $14.7$ s}
    \label{fig:durations}
\end{figure}

\subsection{Impact of the Creativity Program on Conspiratorial Content}
On February 20, 2023, TikTok announced a new Creativity Program for creators, who would be eligible to earn money from their content if they met specific requirements, such as posting videos longer than 1 minute.
These changes took effect starting from May 3 in the U.S. and other countries. 
We already observed in \Cref{sec:RQ1} an overall increment in the quantity of long conspiratorial videos during the period of analysis.
We now investigate whether this effect might be due to the introduction of the Creativity Program.

To this aim, we collect an additional dataset of 10k videos of any duration sampled uniformly at random ($i$) before and ($ii$) after the introduction of the Creativity Program, retrieving the necessary video metadata to extract the length of the content (since the API does not provide this information).
This data is instrumental to measure changes in the whole population of videos, including short ones (duration $< 1$ min).
\Cref{fig:durations} shows the distribution of the durations of collected videos in the two periods.
The number of long videos in the sample increases from N=\num{39} ($0.39\%$) to N=\num{103} ($1.03\%$), probably due to the introduction of the Creativity Program on May 3.
The application of two statistical tests (two-sided Mann-Whitney and Chi-square) confirms that the two distributions are statistically different ($p<0.001$ in both cases).
This analysis indicates that the new Creativity Program influenced platform-wide behavior globally rather than specifically affecting the production of conspiratorial content, as the fraction of conspiracy-related videos remained stable.

\begin{figure}
    \centering
    \includegraphics[width=0.85\linewidth]{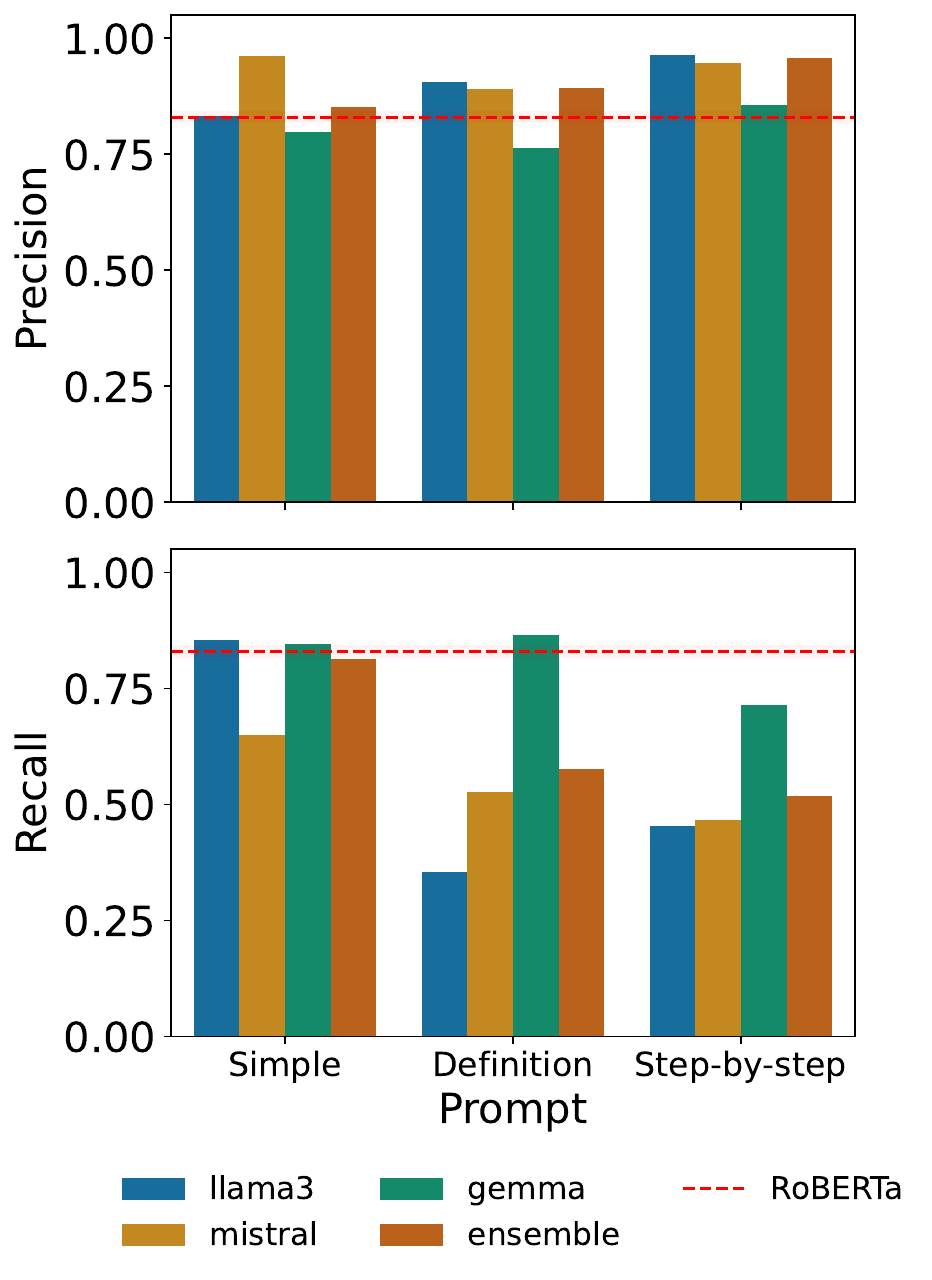}
    \caption{Aggregated results on the positive class (conspiracy videos) of the balanced classification experiment with distant labels (\Cone), for the three prompts and the three models plus the ensemble, across all seeds. The red line is the result of the fine-tuning, 95\% C.I. for RoBERTa.}
    \label{fig:balanced}
\end{figure}

\begin{figure}
\centering
    \includegraphics[width=0.85\linewidth]{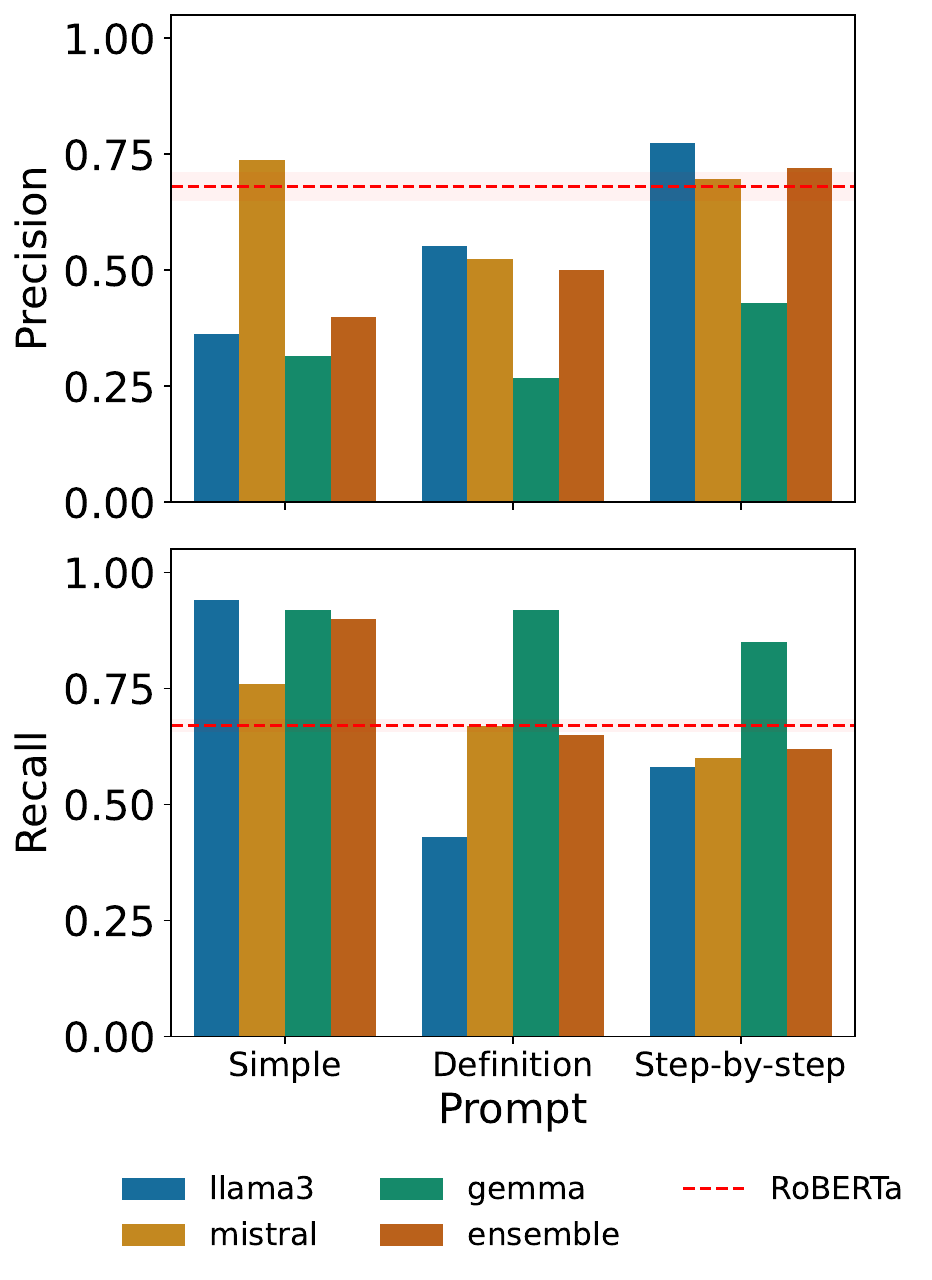}
    \caption{Aggregated results on the positive class (conspiracy videos) of the unbalanced classification experiment with manual labels (\Cthree), for the three prompts and the three models plus the ensemble, across all seeds. The red line is the result of the fine-tuning, 95\% C.I. for RoBERTa.}
    \label{fig:unbalanced}
\end{figure}

\subsection{Detecting Conspiracy Theories with LLMs}
\label{subsec:LLMS}
This section evaluates the performance of different models on identifying conspiratorial videos by measuring their precision ($P$) and recall ($R$).
The last two configurations (\Ctwo and \Cthree), which represent unbalanced settings, are particularly relevant in the context of content moderation, as they highlight the trade-off between false positives (and thus precision) and recall.

As a baseline model, we use RoBERTa base~\cite{DBLP:journals/corr/abs-1907-11692} and fine-tune it on both the balanced (\Cone) ($P = R = 0.83 \pm 0.01 $ on validation) and unbalanced datasets (\Cthree) ($P=0.68 \pm 0.04 $, $R=0.67 \pm 0.01 $ on validation) with a 0.72-0.25 train-test split.
 \Cref{sec:app_roberta} provides further details. 

\Cref{fig:balanced,fig:unbalanced} report the results for the classification task, i.e., detecting the positive (conspiracy) class, specifically for the first (\Cone) and third (\Cthree) configurations.
Additional results for case \Ctwo are included in \Cref{sec:app_classification} and are similar to those for \Cthree.
In both configurations, the three models are surprisingly consistent across all the 25 runs.
There is little to no variation in the classification results across the two datasets and the different prompts.

\spara{Balanced setting (\Cone)}.
Here the different prompts show an interesting pattern (\Cref{fig:balanced}).
The best-performing model, precision-wise, is Llama3 with an almost perfect score (\num{0.96}), with the Step-by-step prompt.
Most models performed better (+1 to +12 pp.) than the baseline, except for Gemma with the Simple and Definition prompts.
Overall, the Step-by-step prompt results in the highest precision, although the difference with the Simple prompt is relatively small.
We see instead an opposite tendency for what concerns the recall: the Simple prompt is the best performing overall, with the highest score of \num{0.87} by Gemma with the Definition prompt.
This last configuration together with Gemma and LLama3 with the Simple prompt are the only ones that perform better than the baseline in terms of recall, while the other models perform significantly worse than RoBERTa (-1 to -49 pp.).

\spara{Unbalanced setting with manual labels (\Cthree)}.
Here we observe a significant decrease in overall performance (\Cref{fig:unbalanced}).
However, the trends identified in the previous configuration remain valid.
The model with the highest precision is again Llama3 with a score of almost \num{0.77} with the Step-by-step prompt.
The same model is the one with the highest recall (\num{0.94}), although with the Simple prompt.
Unfortunately, most of the models perform significantly worse than the baseline in terms of recall (-1 to -49 pp.).
On this dataset, the best overall tradeoff is given by Mistral with the Simple prompt, as it improves over the baseline in both metrics.

\begin{figure*}[t]
\centering
    \includegraphics[width=0.75\linewidth]{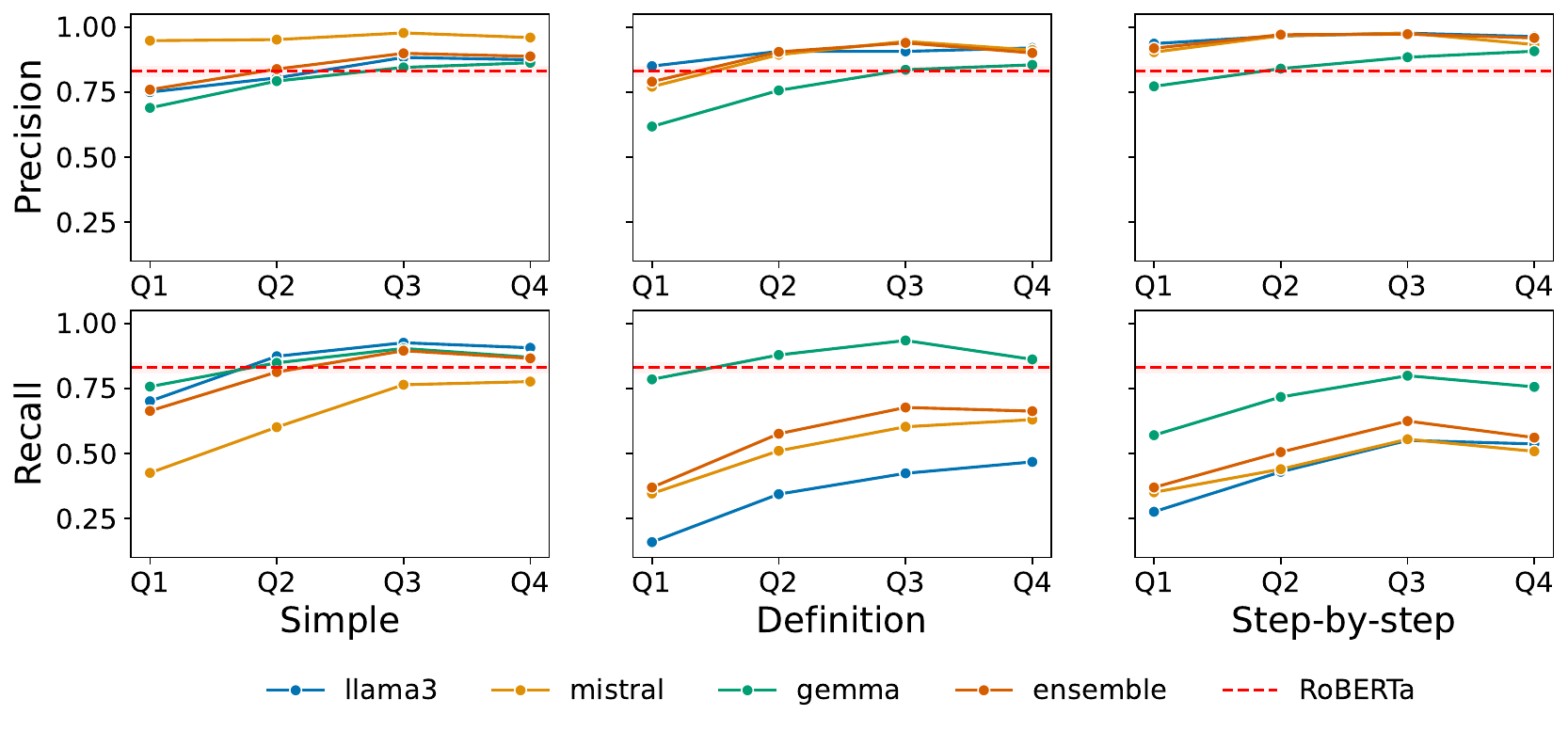}
    \caption{Aggregated results on the positive class (conspiracy videos) of the balanced classification experiment with distant labels (\Cone), for the three prompts and the three models plus the ensemble, across all seeds, split by quartiles on the distribution of the length of the transcriptions in words. The red line is the result of the fine-tuning, 95\% C.I. for RoBERTa. Q1 = $210$ words, Q2 = $325$ words, Q3 = $472$ words, Q4  = $1919$ words.}
    \label{fig:quartiles_results}
\end{figure*}

\spara{Effect of text length.}
Finally, we investigate the relation between classification performance and text length.
\Cref{fig:quartiles_results} charts the classification results for case \Cone for the different quartiles of the text length distribution.
In these plots, we are more interested in the trends rather than the absolute values of the points of the curve (which is better shown in \Cref{fig:balanced,fig:unbalanced}).
Considering precision, the longer the transcript, the higher the metric.
This trend is shared among all the models we consider.
A similar behavior is present for recall, where the metric increases steadily for longer texts, with even a steeper slope than precision.
However, the improvement stops or even reverts when going from \textbf{Q3} to \textbf{Q4}.
For the negative class, RoBERTa is the most precise model in case \Cone, while in \Ctwo and \Cthree LLama3 and Gemma had almost perfect precision (0.99) while being slightly lower in recall (0.98).
We report the full results of the classification process for the negative (not conspiracy) class and additional results for \Cthree in \Cref{sec:app_classification}.

To conclude, note that the ensemble model proves to be a balanced choice, with a better trade-off between precision and recall, but at cost of increased deployment complexity and runtime, since each item needs to be labeled by each of the three different models.
In \Cref{sec:error_anl}, we present a qualitative error analysis, focusing on videos that were unanimously misclassified by all models.
Specifically, we provide summaries for two false positives and two false negatives, along with interpretations of the possible errors.


    

\section{Discussion and Conclusion}
\label{sec:disc}
We collected a representative random sample of approximately 1.5 million videos posted on TikTok in the U.S. via the official Research API.
From this dataset, we identified relevant hashtags associated with prominent and widely known conspiracy theories, such as \emph{illuminati} and \emph{chemtrails}.
Through a combination of hashtag enrichment and manual validation, we identified over a thousand conspiracy-related videos and established a lower bound on the total number of conspiracy theory videos available on the platform.
We then analyzed the correlation between the introduction of TikTok's Creativity Program and the overall increase in the duration of videos posted by users, finding that while conspiracy videos became longer after May 3, 2023 (the starting date for the program), so did non-conspiracy ones.
We investigated the capabilities of open-weight and widely available Large Language Models to classify conspiracy videos given their content in textual format, i.e., the transcript of the speech.
We explored three classification pipelines that represent different scenarios developers and regulators might encounter.
First, we employed a hashtag-based distant supervision labeling approach, which eliminates the need for manual video inspection.
Next, we replicated this pipeline in a more realistic setting, where conspiracy videos constitute a minority, highlighting the impact of class imbalance.
Finally, we refined the pipeline by incorporating manual labeling of the collected conspiracy videos to enhance classification accuracy.

Both classical methods (RoBERTa) and LLMs have their tradeoffs.
Zero-shot prompting removes the cost of training, thus needing only the computational power for inference tasks.
More importantly, it obviates the need to label examples (whether manually or via heuristics such as distant supervision).
Instead, fine-tuning RoBERTa requires the computational and storage capabilities to handle the training and validation of the model.
Still, the inference via LLMs is computationally more expensive than with RoBERTa.
These results should inform decision-making strategies when faced with the deployment of similar systems, e.g., to decide how much effort to allocate to data labeling.
Additionally, the experimental settings we used produced results that are in line with other work in the literature~\citep{atreja2024prompt,ziems2024can}, especially for what concerns the use of definitions and explanations in the prompts and the correlation with increased performance.
We underline that in our setup some configurations surpass the baseline state-of-the-art models, which is a result in contrast with current literature~\citep{plaza-del-arco-etal-2024-wisdom}.
The findings and methodologies presented in this paper have significant implications for both researchers and social media platforms. 
By identifying conspiracy theory videos, our framework can aid in developing more effective moderation strategies and policies to mitigate the diffusion of these harmful contents.
Moreover, the insights gained from analyzing the impact of TikTok's Creativity Program offer valuable perspectives on how platform policies influence content creation trends.
Lastly, LLMs' effectiveness in automating labeling highlights their potential for scalable, data-driven content moderation.
\clearpage

\section*{Limitations}
\label{sec:limit}
As with any empirical work, this study is subject to some limitations.
First, we rely on the TikTok Research API, which functions as an opaque system.
As we lack control over its internal mechanisms, our data may be subject to biases or misrepresentations. 
Additionally, our estimates depend on the initial seed hashtags used in the enrichment process, meaning that different seeds could yield different results.
Our analysis of the Creativity Program's effect is correlational, as the possible presence of hidden confounders prevents us from making causal inferences.
While we provide a detailed and reproducible setup when utilizing LLMs, some degree of uncertainty inherent to these models remains unavoidable.
Also, the models we chose to analyze can have weaker performance compared to larger versions.
In the trade-off between performance and reproducibility together with computational requirements, we chose to prioritize the latter as it allowed us to study the possibility of using these smaller and less costly models to approach the task of content analysis.
Finally, we do not distinguish between the various conspiracy theories present within our collected sample.

Future research should critically examine potential biases in the data provided by the TikTok Research API and compare it with datasets collected directly from the platform.
Enhancing the current work could involve incorporating multimodal data, such as video keyframes, into the classification pipeline. 
This approach may improve the robustness of the analysis.
Similarly, extending the study to other online media platforms and evaluating newly released and bigger models would broaden its scope and applicability.
Additionally, employing Retrieval-Augmented Generation (RAG) to provide contextual information in prompts for LLMs could improve the model's comprehension and accuracy.
Further investigation should also focus on a more detailed quantitative analysis of the behavioral patterns of conspiracy content creators, tracking how these behaviors evolve over time.
Such studies could shed light on the dynamics of conspiracy theory propagation across platforms.

\section*{Ethical considerations}
 TikTok users have explicitly agreed to the Terms of Service, which include the acknowledgment and approval of the transfer of personal data through the API.\footnote{\href{https://www.tiktok.com/legal/page/eea/privacy-policy/en}{TikTok privacy policy} accessed on 12/01}
Throughout the data collection process, we ensured the anonymization of user data to protect individual privacy. 
All results presented are aggregated, i.e., they reflect collective trends rather than identifiable individual behaviors. 
Moreover, no personal information is given as input to the Large Language Models we employed. 
As requested by the Research API Terms of Service, we have implemented robust data security measures to prevent unauthorized access and ensure that all data handling procedures align with current best practices in data protection.
We cannot make our full dataset public, due to the TikTok API usage policies.
We provide access to the code to replicate our results and the dehydrated datasets (Video ID - Label) in the repository associated with the paper.\footnote{Anonymyzed repository: \url{https://anonymous.4open.science/r/ct_tt-FC7E}}
Additionally, while our study highlights the potential of LLMs for detecting conspiratorial content, their current limitations also present an ethical concern, as adversaries may exploit these weaknesses to evade moderation techniques. The models we employed are free to use and attribute ownership of the output to the user (Mistral:Apache2.0\footnote{\href{https://ollama.com/library/mistral/blobs/43070e2d4e53}{Mistral License}}, Llama and Gemma have proprietary custom licenses.\footnote{\href{https://ollama.com/library/gemma2/blobs/097a36493f71}{Gemma License}},\footnote{\href{https://ollama.com/library/llama3/blobs/4fa551d4f938}{Llama License}})

\bibliography{biblio}

\begin{thebibliography}{42}
\providecommand{\natexlab}[1]{#1}

\bibitem[{A{\"\i}meur et~al.(2023)A{\"\i}meur, Amri, and Brassard}]{aimeur2023fake}
Esma A{\"\i}meur, Sabrine Amri, and Gilles Brassard. 2023.
\newblock Fake news, disinformation and misinformation in social media: a review.
\newblock \emph{Social Network Analysis and Mining}, 13(1):30.

\bibitem[{Atreja et~al.(2024)Atreja, Ashkinaze, Li, Mendelsohn, and Hemphill}]{atreja2024prompt}
Shubham Atreja, Joshua Ashkinaze, Lingyao Li, Julia Mendelsohn, and Libby Hemphill. 2024.
\newblock Prompt design matters for computational social science tasks but in unpredictable ways.
\newblock \emph{arXiv preprint arXiv:2406.11980}.

\bibitem[{Augenstein et~al.(2023)Augenstein, Baldwin, Cha, Chakraborty, Ciampaglia, Corney, DiResta, Ferrara, Hale, Halevy et~al.}]{augenstein2023factuality}
Isabelle Augenstein, Timothy Baldwin, Meeyoung Cha, Tanmoy Chakraborty, Giovanni~Luca Ciampaglia, David Corney, Renee DiResta, Emilio Ferrara, Scott Hale, Alon Halevy, et~al. 2023.
\newblock Factuality challenges in the era of large language models.
\newblock \emph{arXiv preprint arXiv:2310.05189}.

\bibitem[{Basch et~al.(2021)Basch, Meleo-Erwin, Fera, Jaime, and Basch}]{basch2021global}
Corey~H Basch, Zoe Meleo-Erwin, Joseph Fera, Christie Jaime, and Charles~E Basch. 2021.
\newblock A global pandemic in the time of viral memes: Covid-19 vaccine misinformation and disinformation on tiktok.
\newblock \emph{Human vaccines \& immunotherapeutics}, 17(8):2373--2377.

\bibitem[{Byford(2011)}]{byford_conspiracy_2011}
J.~Byford. 2011.
\newblock \emph{Conspiracy {Theories}: {A} {Critical} {Introduction}}.
\newblock Springer.
\newblock Google-Books-ID: m5Er9ELOwQkC.

\bibitem[{Chaudhari and Pawar(2022)}]{chaudhari_systematic_2022}
Deptii~D. Chaudhari and Ambika~V. Pawar. 2022.
\newblock \href {https://doi.org/10.4018/JITR.299384} {A {Systematic} {Comparison} of {Machine} {Learning} and {NLP} {Techniques} to {Unveil} {Propaganda} in {Social} {Media}:}.
\newblock \emph{Journal of Information Technology Research}, 15(1):1--14.
\newblock Number: 1.

\bibitem[{Cohen(1960)}]{cohen1960coefficient}
Jacob Cohen. 1960.
\newblock A coefficient of agreement for nominal scales.
\newblock \emph{Educational and psychological measurement}, 20(1):37--46.

\bibitem[{Corso et~al.(2024)Corso, Pierri, and De~Francisci~Morales}]{corso2024we}
Francesco Corso, Francesco Pierri, and Gianmarco De~Francisci~Morales. 2024.
\newblock What we can learn from tiktok through its research api.
\newblock In \emph{Companion Publication of the 16th ACM Web Science Conference}, pages 110--114.

\bibitem[{Diab et~al.(2024)Diab, Nefriana, and Lin}]{diab2024classifying}
Ahmad Diab, Rr. Nefriana, and Yu-Ru Lin. 2024.
\newblock \href {https://doi.org/10.1609/icwsm.v18i1.31318} {Classifying conspiratorial narratives at scale: False alarms and erroneous connections}.
\newblock \emph{Proceedings of the International AAAI Conference on Web and Social Media}, 18(1):340--353.

\bibitem[{Douglas and Sutton(2023)}]{douglas_what_2023}
Karen~M. Douglas and Robbie~M. Sutton. 2023.
\newblock \href {https://doi.org/10.1146/annurev-psych-032420-031329} {What {Are} {Conspiracy} {Theories}? {A} {Definitional} {Approach} to {Their} {Correlates}, {Consequences}, and {Communication}}.
\newblock \emph{Annual Review of Psychology}, 74(1):271--298.
\newblock \_eprint: https://doi.org/10.1146/annurev-psych-032420-031329.

\bibitem[{Douglas et~al.(2019)Douglas, Uscinski, Sutton, Cichocka, Nefes, Ang, and Deravi}]{douglas_understanding_2019}
Karen~M. Douglas, Joseph~E. Uscinski, Robbie~M. Sutton, Aleksandra Cichocka, Turkay Nefes, Chee~Siang Ang, and Farzin Deravi. 2019.
\newblock \href {https://doi.org/10.1111/pops.12568} {Understanding {Conspiracy} {Theories}}.
\newblock \emph{Political Psychology}, 40(S1):3--35.
\newblock \_eprint: https://onlinelibrary.wiley.com/doi/pdf/10.1111/pops.12568.

\bibitem[{Dubey et~al.(2024)Dubey, Jauhri, Pandey, Kadian, Al-Dahle, Letman, Mathur, Schelten, Yang, Fan et~al.}]{dubey2024llama}
Abhimanyu Dubey, Abhinav Jauhri, Abhinav Pandey, Abhishek Kadian, Ahmad Al-Dahle, Aiesha Letman, Akhil Mathur, Alan Schelten, Amy Yang, Angela Fan, et~al. 2024.
\newblock The llama 3 herd of models.
\newblock \emph{arXiv preprint arXiv:2407.21783}.

\bibitem[{Enders et~al.(2023)Enders, Uscinski, Seelig, Klofstad, Wuchty, Funchion, Murthi, Premaratne, and Stoler}]{enders_relationship_2023}
Adam~M. Enders, Joseph~E. Uscinski, Michelle~I. Seelig, Casey~A. Klofstad, Stefan Wuchty, John~R. Funchion, Manohar~N. Murthi, Kamal Premaratne, and Justin Stoler. 2023.
\newblock \href {https://doi.org/10.1007/s11109-021-09734-6} {The {Relationship} {Between} {Social} {Media} {Use} and {Beliefs} in {Conspiracy} {Theories} and {Misinformation}}.
\newblock \emph{Political Behavior}, 45(2):781--804.

\bibitem[{Fong et~al.(2021)Fong, Roozenbeek, Goldwert, Rathje, and Van Der~Linden}]{fong2021language}
Amos Fong, Jon Roozenbeek, Danielle Goldwert, Steven Rathje, and Sander Van Der~Linden. 2021.
\newblock The language of conspiracy: A psychological analysis of speech used by conspiracy theorists and their followers on twitter.
\newblock \emph{Group Processes \& Intergroup Relations}, 24(4):606--623.

\bibitem[{Garimella et~al.(2017)Garimella, Morales, Gionis, and Mathioudakis}]{garimella_quantifying_2017}
Kiran Garimella, Gianmarco De~Francisci Morales, Aristides Gionis, and Michael Mathioudakis. 2017.
\newblock \href {http://arxiv.org/abs/1507.05224} {Quantifying {Controversy} in {Social} {Media}}.
\newblock \emph{arXiv preprint}.
\newblock ArXiv:1507.05224 [cs].

\bibitem[{Golbeck et~al.(2018)Golbeck, Mauriello, Auxier, Bhanushali, Bonk, Bouzaghrane, Buntain, Chanduka, Cheakalos, Everett, Falak, Gieringer, Graney, Hoffman, Huth, Ma, Jha, Khan, Kori, Lewis, Mirano, Mohn~IV, Mussenden, Nelson, Mcwillie, Pant, Shetye, Shrestha, Steinheimer, Subramanian, and Visnansky}]{golbeck_fake_2018}
Jennifer Golbeck, Matthew Mauriello, Brooke Auxier, Keval~H. Bhanushali, Christopher Bonk, Mohamed~Amine Bouzaghrane, Cody Buntain, Riya Chanduka, Paul Cheakalos, Jennine~B. Everett, Waleed Falak, Carl Gieringer, Jack Graney, Kelly~M. Hoffman, Lindsay Huth, Zhenya Ma, Mayanka Jha, Misbah Khan, Varsha Kori, Elo Lewis, George Mirano, William~T. Mohn~IV, Sean Mussenden, Tammie~M. Nelson, Sean Mcwillie, Akshat Pant, Priya Shetye, Rusha Shrestha, Alexandra Steinheimer, Aditya Subramanian, and Gina Visnansky. 2018.
\newblock \href {https://doi.org/10.1145/3201064.3201100} {Fake {News} vs {Satire}: {A} {Dataset} and {Analysis}}.
\newblock In \emph{Proceedings of the 10th {ACM} {Conference} on {Web} {Science}}, {WebSci} '18, pages 17--21, New York, NY, USA. Association for Computing Machinery.

\bibitem[{Good(1953)}]{good1953population}
Irving~J Good. 1953.
\newblock The population frequencies of species and the estimation of population parameters.
\newblock \emph{Biometrika}, 40(3-4):237--264.

\bibitem[{Hadgu et~al.(2013)Hadgu, Garimella, and Weber}]{hadgu2013political}
Asmelash~Teka Hadgu, Kiran Garimella, and Ingmar Weber. 2013.
\newblock Political hashtag hijacking in the us.
\newblock In \emph{Proceedings of the 22nd international conference on world wide web}, pages 55--56.

\bibitem[{Jiang et~al.(2023)Jiang, Sablayrolles, Mensch, Bamford, Chaplot, Casas, Bressand, Lengyel, Lample, Saulnier et~al.}]{jiang2023mistral}
Albert~Q Jiang, Alexandre Sablayrolles, Arthur Mensch, Chris Bamford, Devendra~Singh Chaplot, Diego de~las Casas, Florian Bressand, Gianna Lengyel, Guillaume Lample, Lucile Saulnier, et~al. 2023.
\newblock Mistral 7b.
\newblock \emph{arXiv preprint arXiv:2310.06825}.

\bibitem[{Liu et~al.(2019)Liu, Ott, Goyal, Du, Joshi, Chen, Levy, Lewis, Zettlemoyer, and Stoyanov}]{DBLP:journals/corr/abs-1907-11692}
Yinhan Liu, Myle Ott, Naman Goyal, Jingfei Du, Mandar Joshi, Danqi Chen, Omer Levy, Mike Lewis, Luke Zettlemoyer, and Veselin Stoyanov. 2019.
\newblock \href {https://arxiv.org/abs/1907.11692} {Roberta: {A} robustly optimized {BERT} pretraining approach}.
\newblock \emph{CoRR}, abs/1907.11692.

\bibitem[{Matamoros-Fern{\'a}ndez and Farkas(2021)}]{matamoros2021racism}
Ariadna Matamoros-Fern{\'a}ndez and Johan Farkas. 2021.
\newblock Racism, hate speech, and social media: A systematic review and critique.
\newblock \emph{Television \& new media}, 22(2):205--224.

\bibitem[{Meuer et~al.(2023)Meuer, Oeberst, and Imhoff}]{meuer_how_2023}
Marcel Meuer, Aileen Oeberst, and Roland Imhoff. 2023.
\newblock \href {https://doi.org/10.1002/ejsp.2903} {How do conspiratorial explanations differ from non-conspiratorial explanations? {A} content analysis of real-world online articles}.
\newblock \emph{European Journal of Social Psychology}, 53(2):288--306.
\newblock Number: 2 \_eprint: https://onlinelibrary.wiley.com/doi/pdf/10.1002/ejsp.2903.

\bibitem[{Miani et~al.(2021)Miani, Hills, and Bangerter}]{alessandro_miani_loco_2021}
Alessandro Miani, Thomas~T. Hills, and Adrian Bangerter. 2021.
\newblock \href {https://doi.org/10.3758/s13428-021-01698-z} {{LOCO}: {The} 88-million-word language of conspiracy corpus}.
\newblock \emph{Behavior Research Methods}, pages 1--24.

\bibitem[{Palmer et~al.(2024)Palmer, Smith, and Spirling}]{palmer2024using}
Alexis Palmer, Noah~A Smith, and Arthur Spirling. 2024.
\newblock Using proprietary language models in academic research requires explicit justification.
\newblock \emph{Nature Computational Science}, 4(1):2--3.

\bibitem[{Phillips et~al.(2022)Phillips, Ng, and Carley}]{phillips_hoaxes_2022}
Samantha~C. Phillips, Lynnette Hui~Xian Ng, and Kathleen~M. Carley. 2022.
\newblock \href {https://doi.org/10.1145/3487553.3524665} {Hoaxes and {Hidden} agendas: {A} {Twitter} {Conspiracy} {Theory} {Dataset}: {Data} {Paper}}.
\newblock In \emph{Companion {Proceedings} of the {Web} {Conference} 2022}, {WWW} '22, pages 876--880, New York, NY, USA. Association for Computing Machinery.

\bibitem[{Plaza-del Arco et~al.(2024)Plaza-del Arco, Nozza, and Hovy}]{plaza-del-arco-etal-2024-wisdom}
Flor~Miriam Plaza-del Arco, Debora Nozza, and Dirk Hovy. 2024.
\newblock \href {https://aclanthology.org/2024.nlperspectives-1.2/} {Wisdom of instruction-tuned language model crowds. exploring model label variation}.
\newblock In \emph{Proceedings of the 3rd Workshop on Perspectivist Approaches to NLP (NLPerspectives) @ LREC-COLING 2024}, pages 19--30, Torino, Italia. ELRA and ICCL.

\bibitem[{Pogorelov et~al.(2021)Pogorelov, Schroeder, Filkukov{\'a}, Brenner, and Langguth}]{pogorelov_wico_2021}
Konstantin Pogorelov, Daniel~Thilo Schroeder, Petra Filkukov{\'a}, Stefan Brenner, and Johannes Langguth. 2021.
\newblock \href {https://doi.org/10.1145/3472720.3483617} {{WICO} {Text}: {A} {Labeled} {Dataset} of {Conspiracy} {Theory} and {5G}-{Corona} {Misinformation} {Tweets}}.
\newblock In \emph{Proceedings of the 2021 {Workshop} on {Open} {Challenges} in {Online} {Social} {Networks}}, {OASIS} '21, pages 21--25, New York, NY, USA. Association for Computing Machinery.

\bibitem[{Quaranto(2022)}]{quaranto2022dog}
Anne Quaranto. 2022.
\newblock Dog whistles, covertly coded speech, and the practices that enable them.
\newblock \emph{Synthese}, 200(4):330.

\bibitem[{Radford et~al.(2023)Radford, Kim, Xu, Brockman, McLeavey, and Sutskever}]{radford2023robust}
Alec Radford, Jong~Wook Kim, Tao Xu, Greg Brockman, Christine McLeavey, and Ilya Sutskever. 2023.
\newblock Robust speech recognition via large-scale weak supervision.
\newblock In \emph{International Conference on Machine Learning}, pages 28492--28518. PMLR.

\bibitem[{Saba(2023)}]{saba2023stochastic}
Walid~S Saba. 2023.
\newblock Stochastic llms do not understand language: towards symbolic, explainable and ontologically based llms.
\newblock In \emph{International Conference on Conceptual Modeling}, pages 3--19. Springer.

\bibitem[{Sclar et~al.(2023)Sclar, Choi, Tsvetkov, and Suhr}]{sclar2023quantifying}
Melanie Sclar, Yejin Choi, Yulia Tsvetkov, and Alane Suhr. 2023.
\newblock Quantifying language models' sensitivity to spurious features in prompt design or: How i learned to start worrying about prompt formatting.
\newblock \emph{arXiv preprint arXiv:2310.11324}.

\bibitem[{Spohr(2017)}]{spohr2017fake}
Dominic Spohr. 2017.
\newblock Fake news and ideological polarization: Filter bubbles and selective exposure on social media.
\newblock \emph{Business information review}, 34(3):150--160.

\bibitem[{Sutton and Douglas(2020)}]{sutton_conspiracy_2020}
Robbie~M Sutton and Karen~M Douglas. 2020.
\newblock \href {https://doi.org/10.1016/j.cobeha.2020.02.015} {Conspiracy theories and the conspiracy mindset: implications for political ideology}.
\newblock \emph{Current Opinion in Behavioral Sciences}, 34:118--122.

\bibitem[{Team et~al.(2024)Team, Mesnard, Hardin, Dadashi, Bhupatiraju, Pathak, Sifre, Rivi{\`e}re, Kale, Love et~al.}]{team2024gemma}
Gemma Team, Thomas Mesnard, Cassidy Hardin, Robert Dadashi, Surya Bhupatiraju, Shreya Pathak, Laurent Sifre, Morgane Rivi{\`e}re, Mihir~Sanjay Kale, Juliette Love, et~al. 2024.
\newblock Gemma: Open models based on gemini research and technology.
\newblock \emph{arXiv preprint arXiv:2403.08295}.

\bibitem[{T{\"o}rnberg(2024)}]{tornberg2024best}
Petter T{\"o}rnberg. 2024.
\newblock Best practices for text annotation with large language models.
\newblock \emph{arXiv preprint arXiv:2402.05129}.

\bibitem[{Uscinski et~al.(2017)Uscinski, Douglas, and Lewandowsky}]{uscinski2017climate}
Joseph~E. Uscinski, Karen Douglas, and Stephan Lewandowsky. 2017.
\newblock \href {https://doi.org/10.1093/acrefore/9780190228620.013.328} {Climate {Change} {Conspiracy} {Theories}}.
\newblock In \emph{Oxford {Research} {Encyclopedia} of {Climate} {Science}}. Oxford University Press.

\bibitem[{Wang et~al.(2003)Wang, Acero, and Chelba}]{wang2003word}
Ye-Yi Wang, Alex Acero, and Ciprian Chelba. 2003.
\newblock Is word error rate a good indicator for spoken language understanding accuracy.
\newblock In \emph{2003 IEEE workshop on automatic speech recognition and understanding (IEEE Cat. No. 03EX721)}, pages 577--582. IEEE.

\bibitem[{Wei et~al.(2022)Wei, Wang, Schuurmans, Bosma, Xia, Chi, Le, Zhou et~al.}]{wei2022chain}
Jason Wei, Xuezhi Wang, Dale Schuurmans, Maarten Bosma, Fei Xia, Ed~Chi, Quoc~V Le, Denny Zhou, et~al. 2022.
\newblock Chain-of-thought prompting elicits reasoning in large language models.
\newblock \emph{Advances in neural information processing systems}, 35:24824--24837.

\bibitem[{Weimann and Masri(2023)}]{weimann2023research}
Gabriel Weimann and Natalie Masri. 2023.
\newblock Research note: Spreading hate on tiktok.
\newblock \emph{Studies in conflict \& terrorism}, 46(5):752--765.

\bibitem[{Zeng and Kaye(2022)}]{zeng2022content}
Jing Zeng and D~Bondy~Valdovinos Kaye. 2022.
\newblock From content moderation to visibility moderation: A case study of platform governance on tiktok.
\newblock \emph{Policy \& Internet}, 14(1):79--95.

\bibitem[{Zenone and Caulfield(2022)}]{zenone2022using}
Marco Zenone and Timothy Caulfield. 2022.
\newblock Using data from a short video social media platform to identify emergent monkeypox conspiracy theories.
\newblock \emph{JAMA Network Open}, 5(10):e2236993--e2236993.

\bibitem[{Ziems et~al.(2024)Ziems, Held, Shaikh, Chen, Zhang, and Yang}]{ziems2024can}
Caleb Ziems, William Held, Omar Shaikh, Jiaao Chen, Zhehao Zhang, and Diyi Yang. 2024.
\newblock Can large language models transform computational social science?
\newblock \emph{Computational Linguistics}, 50(1):237--291.

\end{thebibliography}
\clearpage
\appendix

\section{Appendix}
\label{sec:appendixA}

\subsection{Seed hashtags}
\label{sec:app_seeds}
\begin{table*}
\caption{Seeds from LOCO used as hashtags for the enrichment process along with a brief description of the theory.}
\label{tab:seeds_and_description}
\footnotesize
\begin{tabular}{ll}

\toprule
\textbf{Seed} & \textbf{Description} \\ \midrule
conspiracy     &   General seed of the topic.          \\ \cmidrule(lr){1-2}
flatearth     &    \makecell[l]{The flat Earth conspiracy theory posits that the Earth is not a sphere but a flat, disc-shaped plane, \\contrary to centuries of scientific evidence and exploration.}    \\ \cmidrule(lr){1-2}
qanon     &       \makecell[l]{The QAnon conspiracy theory alleges that a secret cabal of Satan-worshipping pedophiles,\\ including prominent politicians and celebrities, is controlling the world\\ and that former President Donald Trump is working to dismantle this group. }     \\\cmidrule(lr){1-2}
newworldorder     &  \makecell[l]{The New World Order conspiracy theory asserts that a clandestine group of powerful elites \\is conspiring to establish a single, global government that would control all aspects of life\\ and suppress individual freedoms. }         \\\cmidrule(lr){1-2}
chemtrails     &    \makecell[l]{The chemtrails conspiracy theory claims that the trails left by aircraft in the sky, known as contrails,\\ are actually chemicals being deliberately sprayed for nefarious purposes,\\ such as weather manipulation, population control, or biological warfare. }         \\\cmidrule(lr){1-2}
mindcontrol     & \makecell[l] {The mind control conspiracy theory suggests that governments, corporations, or other powerful entities\\ are using advanced technologies or psychological techniques to manipulate and control the thoughts,\\ behaviors, and perceptions of individuals.}         \\\cmidrule(lr){1-2}
reptilian     &    \makecell[l]  {The reptilian conspiracy theory posits that shape-shifting reptilian aliens, disguised as humans, \\ are controlling the world by infiltrating positions of power in government, media, and industry. }        \\\cmidrule(lr){1-2}
bigfoot   &     \makecell[l]{The Bigfoot conspiracy theory suggests that the existence of Bigfoot, a large, ape-like creature \\said to inhabit remote forests, is being deliberately covered up by governments or other authorities.\\ Proponents believe that evidence of Bigfoot's existence is suppressed to prevent public knowledge }       \\ \cmidrule(lr){1-2}
illuminati    &    \makecell[l]{The Illuminati conspiracy theory asserts that a secret society called the Illuminati is covertly orchestrating \\ global events and manipulating governments, economies, and media to establish a New World Order.}          \\\cmidrule(lr){1-2}
ufo     &      \makecell[l] {The UFO conspiracy theory claims that governments, particularly the U.S. government, are hiding evidence \\ of unidentified flying objects (UFOs) and extraterrestrial life from the public. } \\
\bottomrule
\end{tabular}
\end{table*}

\Cref{tab:seeds_and_description} presents the 10 seed keywords used for the hashtag enrichment process, which include nine terms from LOCO dataset~\cite{alessandro_miani_loco_2021} along with the keyword \emph{conspiracy}, accompanied by brief descriptions.



\subsection{Randomness in LLMs experiments}
\label{sec:app_llms}
We assess the model's consistency by using different seeds during the classification process.
Setting a seed ensures the reproducibility of our results.
Here we list the seeds used in our classification process:\\
\num{0}, \num{4}, \num{7}, \num{8}, \num{10}, \num{12}, \num{17}, \num{25}, \num{26}, \num{33}, \num{37}, \num{42}, \num{49}, \num{50}, \num{57}, \num{69}, \num{73}, \num{100}, \num{123}, \num{420}, \num{444}, \num{666}, \num{777}, \num{1111}, \num{1999}

\subsection{Data Collection Process}
\label{sec:app_data}
\begin{figure}[!t]
    \centering
    \includegraphics[width=1\linewidth]{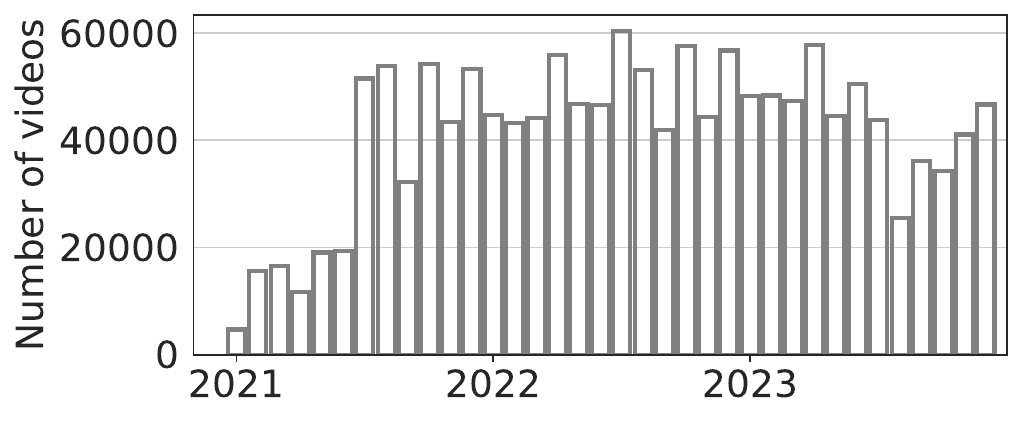}
    \caption{Monthly number of unique long videos (duration $\geq 1$ minute) collected from 2021 to 2023.}
    \label{fig:time_collection}
\end{figure}

\begin{figure}[!t]
    \centering
    \includegraphics[width=1\linewidth]{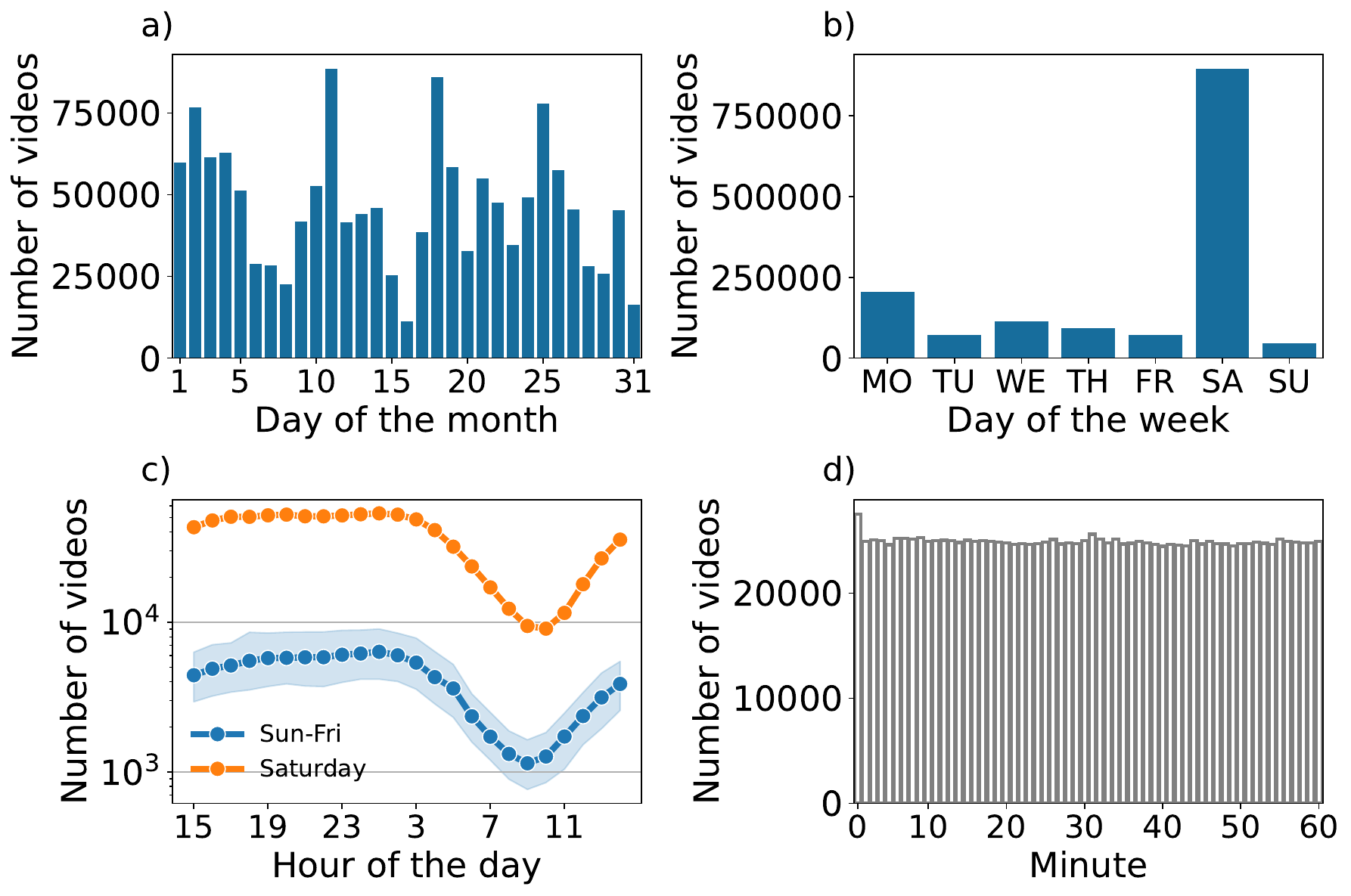}
    \caption{Number of videos posted (a) for each day of the month, (b) for each day of the week, (c) for each hour of the day (UTC), and (d) for each minute of the hour.
    The API shows a clear bias at the daily level, with a non-uniform distribution over different days of the month, and a disproportionate number of videos shared on Saturdays.}
    \label{fig:temporal_bias}
\end{figure}

We collect our data by stratifying samples by week to mitigate issues observed in previous research, particularly the non-uniform distribution of data returned by the random sample endpoint~\cite{corso2024we}.
We focus on collecting long videos, i.e., those with duration $\geq$ 1 minute, as they are expected to provide sufficient contextual information for effective analysis by LLMs.
We perform requests to the \texttt{/video/query} endpoint for each week from January 2021 to December 2023, retaining only videos published in the U.S. and setting the parameter \texttt{is\_random = True}.
\Cref{fig:time_collection} reports the time series of collected unique videos.
There is a visible change in the number of collected videos starting from July 2021, when TikTok started allowing videos up to three minutes in length to be uploaded.
We conduct exploratory analyses of the videos' temporal features to build on previous work and evaluate the effectiveness of our sampling technique.
\Cref{fig:temporal_bias} shows the results of these evaluations.
Sub-figure \textbf{a} shows the number of videos posted for each given day of the month: all days are represented,  with some expected variability.
Sub-figures \textbf{b,c} depict similar behavior to what has been already described, with a majority of the videos created on Saturdays and the normal circadian world rhythm~\cite{corso2024we}.
Finally, sub-figure \textbf{d} reveals that the first minute of the hour presents a slightly higher number of videos, possibly due to the scheduling of videos by creator accounts.

\begin{table*}
\caption{Classification results for the negative class (not conspiracy videos) , for the different LLM configurations with \Cone balanced dataset with distant labels (Left), \Ctwo unbalanced with distant labels (Center), and \Cthree unbalanced with manual labels (Right).
SP = Simple Prompt, DP = Definition Prompt, SBS = Step-by-step prompt.}
\label{tab:class-neg}
\scriptsize
\begin{tabular}{@{}lcccccccccccccccccc@{}}
\toprule
        Configuration & \multicolumn{6}{c}{\textbf{Balanced with distant labels}}              & \multicolumn{6}{c}{\textbf{Unbalanced with distant labels}}         & \multicolumn{6}{c}{\textbf{Unbalanced with manual labels}}          \\
         \cmidrule(lr){2-7} \cmidrule(lr){8-13} \cmidrule(lr){14-19}
Measure   & \multicolumn{3}{c}{Precision} & \multicolumn{3}{c}{Recall}    & \multicolumn{3}{c}{Precision} & \multicolumn{3}{c}{Recall} & \multicolumn{3}{c}{Precision} & \multicolumn{3}{c}{Recall} \\
         \cmidrule(lr){2-4} \cmidrule(lr){5-7} \cmidrule(lr){8-10} \cmidrule(lr){11-13} \cmidrule(lr){14-16} \cmidrule(lr){17-19}
Prompt         & SP       & DP       & SBS     & SP       & DP      & \multicolumn{1}{c}{SBS}     & SP       & DP       & SBS     & SP      & DP      & SBS    & SP       & DP       & SBS     & SP       & DP     & SBS    \\ \cmidrule{1-19}
LLama3   & 0.84     & 0.58     & 0.62    & 0.81     & 0.96     & \multicolumn{1}{c}{0.98}    & 0.98     & 0.92     & 0.98    & 0.81    & 0.96    & \multicolumn{1}{c}{0.98}   & 0.99     & 0.93     & 0.95    & 0.80     & 0.96     & 0.98   \\
Mistral  & 0.72     & 0.63     & 0.62    & 0.97     & 0.81     & \multicolumn{1}{c}{0.90}    & 0.95     & 0.94     & 0.94    & 0.97    & 0.80    & \multicolumn{1}{c}{0.91}   & 0.97     & 0.96     & 0.95    & 0.97     & 0.81     & 0.91   \\
Gemma    & 0.83     & 0.84     & 0.74    & 0.77     & 0.67     & \multicolumn{1}{c}{0.86}    & 0.97     & 0.98     & 0.97    & 0.77    & 0.69    & \multicolumn{1}{c}{0.87}   & 0.99     & 0.99     & 0.98    & 0.76     & 0.66     & 0.86   \\
Ensemble & 0.80     & 0.66     & 0.65    & 0.84     & 0.92    & \multicolumn{1}{c}{0.98}    & 0.97     & 0.93     & 0.94    & 0.84    & 0.93    & \multicolumn{1}{c}{0.98}   & 0.99     & 0.95     & 0.95    & 0.84     & 0.92     & 0.97   \\
RoBERTa  & \multicolumn{3}{c}{0.88 $\pm$ 0.02} & \multicolumn{3}{c}{0.86  $\pm$ 0.03} & \multicolumn{3}{c}{0.95 $\pm$ 0.01}      & \multicolumn{3}{l}{0.97  $\pm$ 0.01}   & \multicolumn{3}{c}{0.95 $\pm$ 0.01}      & \multicolumn{3}{c}{0.97  $\pm$ 0.01}   \\
\bottomrule
\end{tabular}
\end{table*}

\subsection{Fine-tuning RoBERTa }
\label{sec:app_roberta}
As a baseline, we use RoBERTa~\cite{DBLP:journals/corr/abs-1907-11692}, available on HuggingFace.
We adopt the default training pipeline provided by the \texttt{tranformers} library and fine-tune the model by using both balanced and unbalanced datasets with distant labels.
We use a training-test split of 0.75-0.25 for 5 epochs with a learning rate of $\alpha = 2 \times 10^{-5}$ and we instruct the default trainer to keep the best model at the end.

\begin{figure}[!t]
    \Centering
    \includegraphics[width=0.9\linewidth]{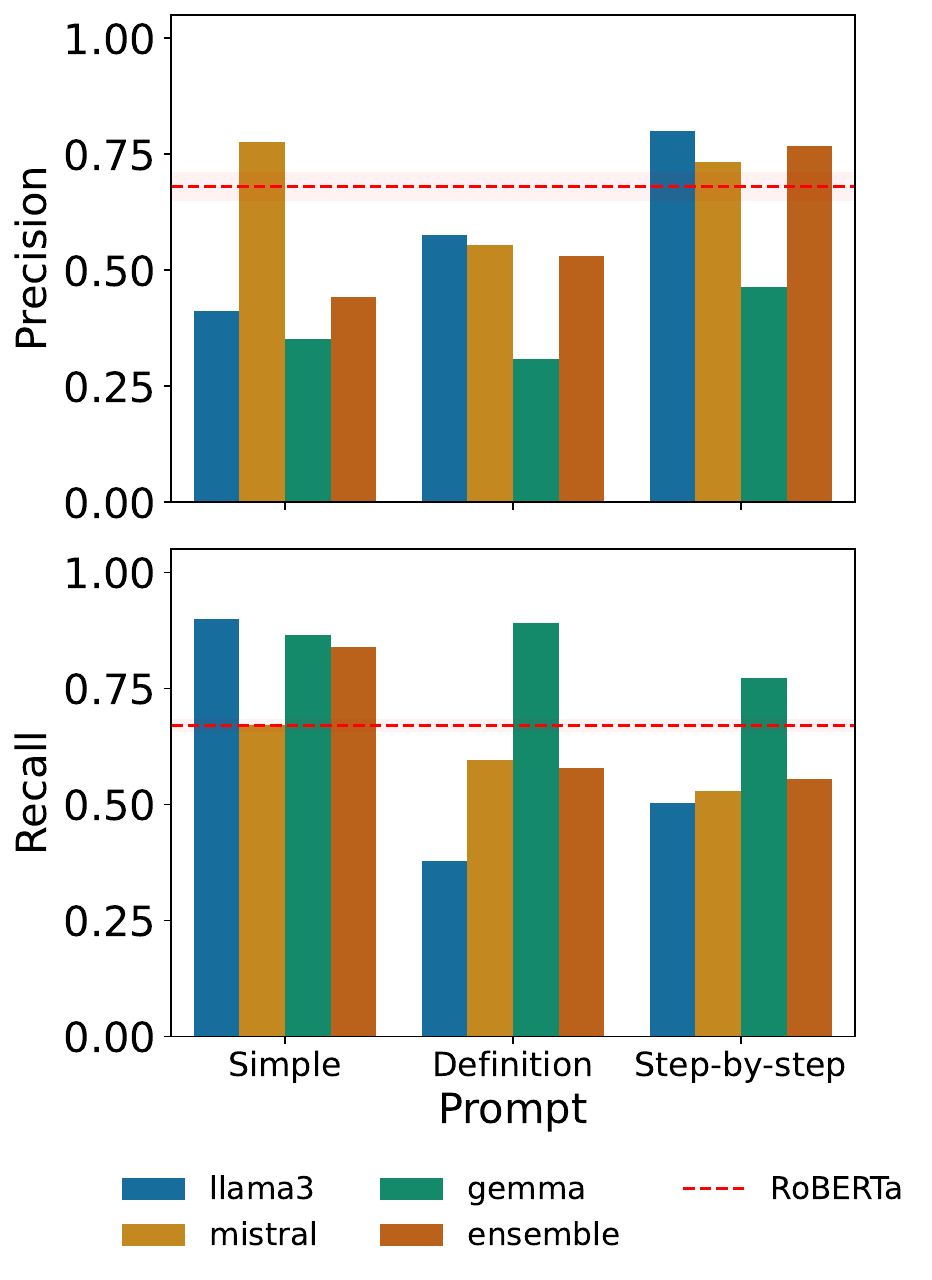}
    \caption{Aggregated results on the positive class of the unbalanced classification experiment with distant labels (\Ctwo), for the three prompts and the three models plus the ensemble, across all seeds. The red line is the result of the fine-tuning, C.I 95\% for RoBERTa (the same used for case \Cthree).}
    \label{fig:distant_unbalanced}
\end{figure}

\subsection{Additional Classification Results}
\label{sec:app_classification}
\begin{figure*}[!t]
\centering
    \includegraphics[width=.8\linewidth]{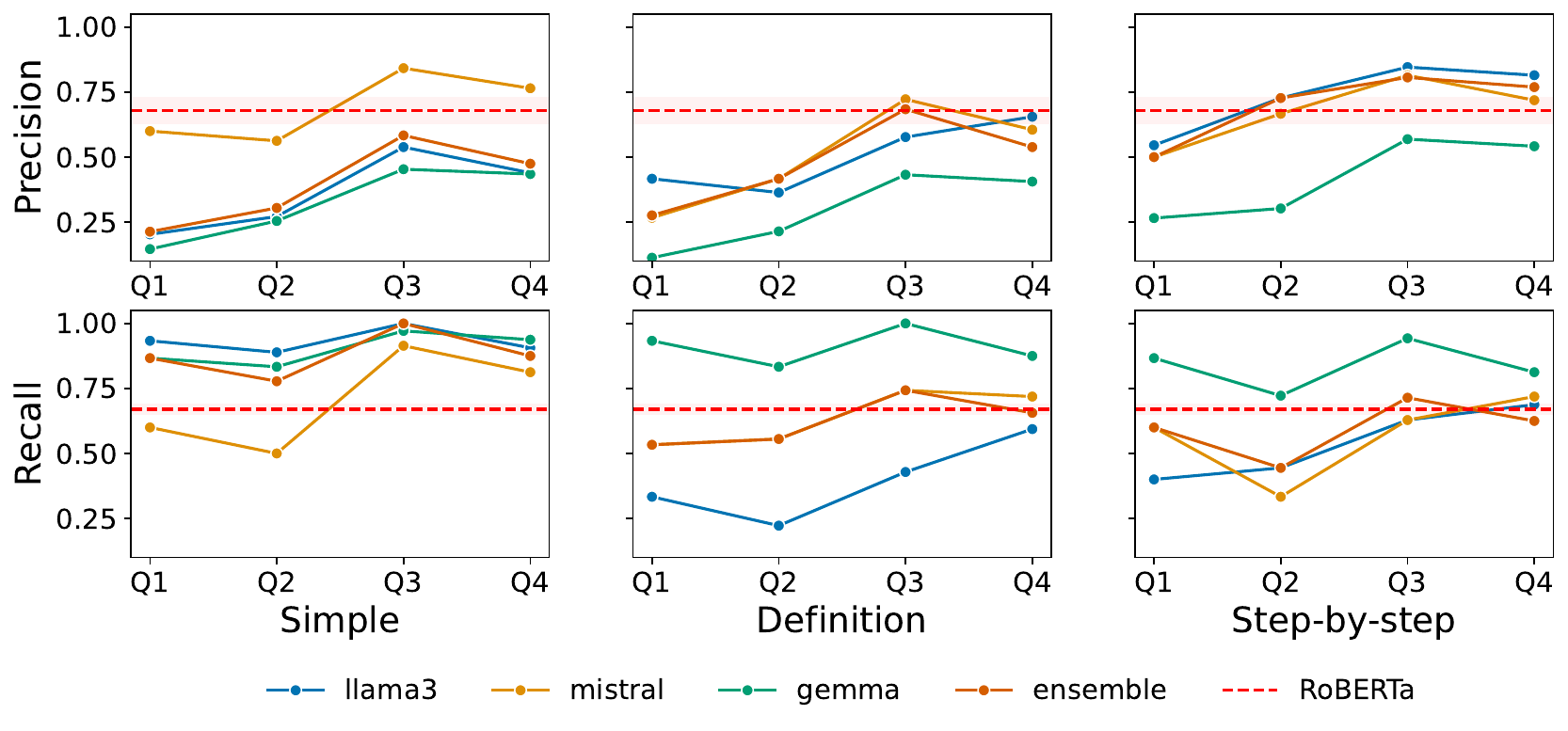}
    \caption{Aggregated results on the positive class (conspiracy videos) of the unbalanced classification experiment with manual labels (\Cthree), for the three prompts and the three models plus the ensemble, across all seeds, split by quartiles on the distribution of the length of the transcriptions in words. The red line is the result of the fine-tuning, 95\% C.I. for RoBERTa.
    Q1 = $210$ words, Q2 = $325$ words, Q3 = $472$ words, Q4  = $1919$ words.}
    \label{fig:quartiles_results_unb}
\end{figure*}

\Cref{fig:distant_unbalanced} reports the trends for Precision and Recall in \Cthree. 
\Cref{tab:class-neg} lists the classification results for the negative class.
\Cref{fig:quartiles_results_unb} shows the classification results for the positive class with the unbalanced dataset and distant labels.
The results are similar to the ones for the configuration \Cone analyzed in \Cref{sec:results}.
Generally, giving longer texts increases the metrics. 
However, in some cases, long texts (\textbf{Q4}) perform worse than slightly shorter ones (\textbf{Q3}) in Precision and Recall.

\subsection{Error Analysis}
\label{sec:error_anl}
We provide examples of errors (\textbf{False Positive}: the models interpreted the transcription as conspiratorial when it is not, \textbf{False Negative}: the transcription mentioned conspiracy theories but the models did not capture it) committed by the models unanimously (all three of them misclassified the item).

First, we show two \textbf{False Negatives} summarizing their \texttt{voice\_to\_text}:
\begin{itemize}
    \item  \textbf{Dome-Earth:} The speaker rejects the flat-earth theory but claims the Earth is under a dome, making references to the Torah and older versions of the Bible that describe the creation of a ``firmament'' supporting their dome theory. They argue that NASA’s images of Earth are Photoshopped and claim NASA admits this.

    \item  \textbf{Big Foot:} The speaker mentions hearing distinct knocking sounds on wood, a behavior often associated with Bigfoot sightings. This knocking occurred in a quiet neighborhood where few people are outside after dark. The speaker links the current experience to a previous instance in the same area where they believe they encountered Bigfoot. Notes that local dogs often bark uncontrollably at night, adding to the sense of something unusual occurring.
\end{itemize}
Both examples explicitly reference a conspiracy theory; however, the models failed to accurately detect it. One possible explanation lies in the tone of the content. In the first example, the rejection of the conspiracy theory (flath-earth) in favour of another theory (dome-earth) may have influenced the model's output, as its sequential processing of the text might have led it to misinterpret the context.
The second example explicitly mentions ``Big Foot'', yet the model appears to have categorized the content as mystery or creepy-themed material rather than identifying it as a conspiracy theory.

We then provide two \textbf{False Positives} examples:

\begin{itemize}
    \item \textbf{God and racism:} The speaker asserts that racism in humans is a character flaw. However, they distinguish this from divine sovereignty, claiming God has the prerogative to show preferential treatment. The argument centers on God's preference for Israel in the Old Testament. One speaker suggests this preference could be interpreted as ``racism,'' while the other counters that it was divine prerogative, not racism.

    \item \textbf{Woke Culture:} The speaker references a previous video about ``fake woke'' culture, criticizing what they perceive as excessive political correctness, particularly on the West Coast. They mention a comedic skit by two comedians, which they believe humorously illustrates their point about the absurdities of extreme ``woke'' ideologies such as White Privilege, Racially Exclusive Spaces and Cultural Appropriation. The skit appears to parody extreme ``woke'' ideologies by exaggerating contradictions, hypocrisy, and rigid social rules.
\end{itemize}

These examples were both wrongly classified as containing conspiracy theories.
In the first example, we can suppose the models were driven to label the sample as positive due to the presence of topics such as ``God'' and ``Israel.''
In the second example instead, the models may fail to capture the comical context of the transcript and instead focus on the literal political meaning of the text and the possible conspiracies associated with it.

We provide a manual interpretation rather than relying on the model-generated explanations, as these are neither reproducible nor reflective of the model's true reasoning process~\cite{saba2023stochastic}.

\subsection{Evaluation of Whisper accuracy}
\label{sec:app_whisper}
Some videos (<5\%) have unintelligible content or spoken content in languages different from English.
We discard these videos from our sample via simple text filtering based on the obtained transcription from the TikTok Research API.
We then manually code the transcriptions of a set of \num{100} videos drawn randomly from our sample.
This process consisted in watching every video of the sample and manually transcribing the content of the audio.

\subsection{Transcription Word Lengths }
\label{sec:app_transcr}
\begin{figure}[!t]
    \centering
    \includegraphics[width=\linewidth]{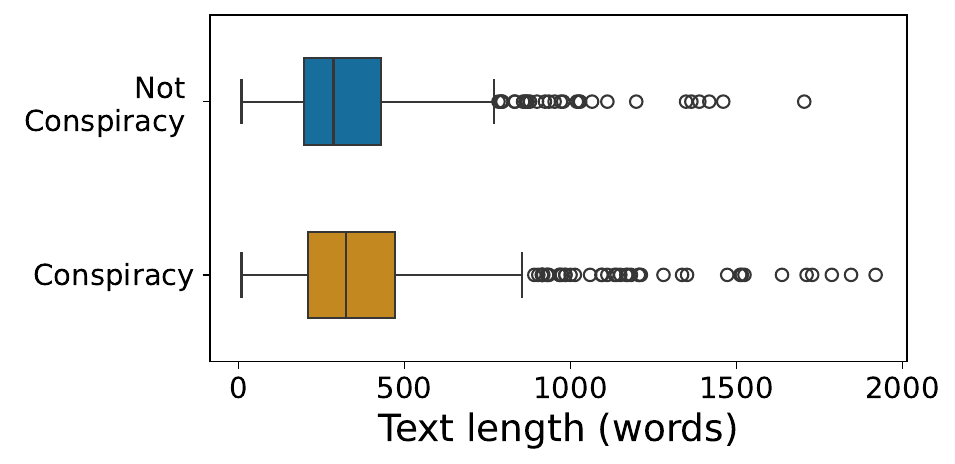}
    \caption{Distributions of the length of the video transcripts for the positive (conspiracy) and negative (not conspiracy) class in the balanced dataset. Medians: Not Conspiracy = \num{286} words, Conspiracy  = \num{325} words}
    \label{fig:durations_text}
\end{figure}

\Cref{fig:durations_text} shows the per-class distributions of the length (in words) of the transcripts extracted by Whisper.
There is a small yet significant statistical difference between the two distributions (two-sided Mann-Whitney, $p<0.001$).
\Cref{subsec:LLMS} investigates the correlation between the length of the transcript and the performance of LLMs.

\begin{figure}[!t]
    \centering
    \includegraphics[width=0.7\linewidth]{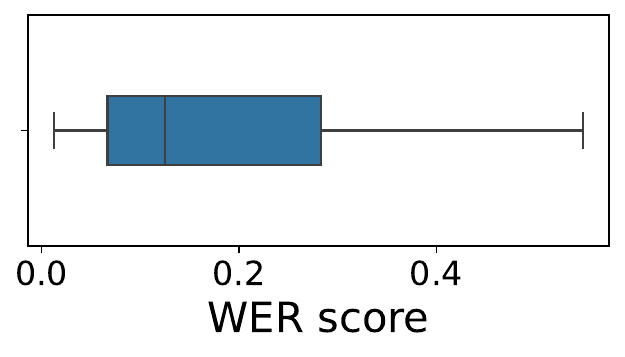}
    \caption{Distribution of WER scores for videos transcribed by Whisper compared to manual coding.}
    \label{fig:wer}
\end{figure}

\section{Definition of Conspiracy Theory}
\label{app:definition}
We use the definition of conspiracy theory provided by \citet{douglas_what_2023} in the \textit{Definition Prompt}.
\textit{`A conspiracy theory is a belief that two or more actors have coordinated in secret to achieve an outcome and that their conspiracy is of public interest but not public knowledge. Conspiracy theories (a) are oppositional, which means they oppose publicly accepted understandings of events; (b) describe malevolent or forbidden acts; (c) ascribe agency to individuals and groups rather than to impersonal or systemic forces; (d) are epistemically risky, meaning that though they are not necessarily false or implausible, taken collectively they are more prone to falsity than other types of belief; and (e) are social constructs that are not merely adopted by individuals but are shared with social objectives in mind, and they have the potential not only to represent and interpret reality but also to fashion new social realities'}.

\end{document}